# Thermodynamics of irreversible particle creation phenomena and its cosmological consequence

Abhik Kumar Sanyal[1], Subhra Debnath[2]

Dept. of Physics, Jangipur College, Jangipur, (Affiliated to University of Kalyani) Murshidabad, India, Pin: 742213

[1] sanyal_ak@yahoo.com

[2] subhra_dbnth@yahoo.com

**Abstract.** The study of particle creation phenomena at the expense of the gravitational field is of great research interest. It might solve the cosmological puzzle singlehandedly, without the need for either dark energy or modified theory of gravity. In the early universe, following graceful exit from inflationary phase, it serves the purpose of reheating the cold universe, which gave way to the hot Big-Bang model. In the late universe, it led to late time cosmic acceleration, without affecting standard Big-Bang-Nucleosynthesis (BBN), Cosmic Microwave Background Radiation (CMBR), or Structure Formation. In this chapter, we briefly review the present status of cosmic evolution, develop the thermodynamics for irreversible particle creation phenomena and study its consequences at the early as well as at the late universe.

## 1   Introduction

Despite Hubble's discovery in 1929, that the universe is expanding, being supported by Friedmann - Lemaître's so called standard model of cosmology, and detection of cosmic microwave background radiation (CMBR) by Penzias and Wilson in 1965, the real birth of modern cosmology took place only after Alan Guth's seminal paper on inflationary scenario in 1981. Since then, general relativists and particle physicists are working hand-in-hand to explore the evolution of the universe from very early stage, till date. However, only after the detection of cosmic microwave background anisotropies (which are the source of the seeds of perturbations required for structure formation), by cosmic background explorer satellite (COBE) in 1992, observational cosmology took birth.

With the advent of advanced satellite based technology, different space-agencies initiated modern observational cosmology. Two more satellites, the WMAP (Wilkinson microwave anisotropy probe) by NASA and the Planck by European space agency, were launched thereafter. These new technologies could explore anisotropies down to angular scales of a few arc-minutes. The experimental data confirmed early Inflationary era of cosmic evolution and thereby placed Inflation as a pre Big-Bang scenario rather than simply a model. Inflation solved many of the problems of the standard model of cosmology, viz. the horizon problem, the flatness problem, the monopole problem etc. Most importantly, it gives rise to the seeds of perturbation, required for structure formation. At the end of Inflation the Universe is reheated giving birth to a thick, hot soup of Plasma – the Big-bang. In the process, the initial singularity has been pushed very close to the Planck's era and Big-Bang is no longer a singularity. The standard model works fairly well thereafter, explaining all the observed phenomena – the nucleosynthesis, formation of microwave background radiation, the structure formation etc. However, SNIa data puzzled cosmologists for last two decades, since it supports present accelerated expansion of the universe. The standard model again has no answer to such observation.

Inflation is usually driven by a scalar field (or more), and the field decays quickly giving way to the standard model. Recently detected Higgs might also be a responsible candidate for inflation. But, to explain late-stage of cosmic evolution again one (or more) exotic scalar field is required. Though such fields have theoretical origin, there is no possibility of detecting such fields, since they interact with nothing but gravity, and therefore are dubbed as dark energy.

Modified Theory of Gravity on the other hand can solve the puzzle without the requirement of such exotic scalars. Indeed, in the very early universe, Einstein's theory is required to be modified by the inclusion of higher-order curvature invariant terms. This is because, General theory of Relativity (GTR) is non-renormalizable, and it is understood that GTR should be realized as the weak field (classical) approximation of a renormalized and unitary quantum theory of gravity, and that too may be in higher dimension. Nevertheless, modification of GTR at the late stage of cosmological evolution, sounds rather artificial, since the curvature invariant terms required to explain late-stage of cosmological evolution, don't arise from any meaningful physical argument.

On the contrary, particle creation phenomena at the cost of gravitational field in the early universe has been studied extensively by Parker and his collaborators (Parker 1968, 1969, 1971; Papastamatiou and Parker 1979) in the last century. Thereafter, Prigogine and his collaborators (Prigogine et al. 1989; Prigogine 1989) treated inflation as a paradigm of irreversible, adiabatic particle creation phenomena, which neither required a scalar field, nor modification of gravity. Such an inflationary model can give way to reheating following collision of created particles. In recent years, Lima and his collaborators (Lima et al. 2008; Steigman et al. 2009) and later Debnath and Sanyal (Debnath and Sanyal 2011) could successfully explain late-stage of cosmic evolution following very slow particle creation rate, in nearly flat Robertson-Walker model. In this chapter, we briefly discuss different attempts made over decades to explain the observed cosmic evolution, since late sixties of last century. Thereafter, we detail the phenomena of particle creation in the early as well as the late stage of cosmic evolution, as an alternative theory.

The scheme of the present chapter runs as follows. In the following section, we briefly discuss the standard model of cosmology. In section 3, we discuss dissipative phenomena, which were initiated before the birth of inflation to explain presently observed isotropy and homogeneity of the universe. In section 4, we discuss the basic essence of Inflationary scenario. In section 5, we present to somewhat detailed the thermodynamics of irreversible particle creation phenomena. The Inflationary paradigm following such phenomena has also been discussed. In section 6, recent observations in connection with accelerated expansion in the late-stage of cosmic evolution, and its possible resolutions following dark energy and modification of gravity have been addressed in brief. In the same section, we explained in the role of slow particle creation in the nearly flat Robertson-Walker metric, to explain the recent observations, in some detail.. Concluding remarks appear in chapter 7.

## 2    The Standard model of cosmology

Standard model of cosmology (see for example (Wienberg 1972; Narlikar 1993; Islam 2002)) is based on cosmological principle, which states that the universe is spatially homogeneous and isotropic on large scales $> 100$Mpc. Such an assumption is described by the Robertson-Walker (RW) metric which is given by

$$ds^2 = -dt^2 + a^2(t)\left[\frac{dr^2}{1-kr^2} + r^2(d\theta^2 + \sin^2\theta \, d\phi^2)\right] \qquad (1)$$

where, $t$ is the cosmic time, $r, \theta, \phi$ are spatial co-ordinates, $a(t)$ is the scale factor of the universe and $k = 0, \pm 1$ is the three space curvature parameter, corresponding to flat, closed and open universe respectively. Now assuming that the universe is filled with perfect fluid, the energy-momentum tensor is given by

$$T_{\mu\nu} = (p + \rho)v_\mu v_\nu + p \, g_{\mu\nu} \qquad (2)$$

where, $v_\mu$ is the 4-velocity vector, $g_{\mu\nu}$ is the metric tensor, $p$ is the isotropic pressure, and $\rho$ is the energy density of the perfect fluid. The dynamics of the space-time geometry is governed by Einstein's equation which relates the geometry of the universe to its energy content as $G_{\mu\nu} = 8\pi G T_{\mu\nu}$, where, $G$ is universal gravitational constant and $G_{\mu\nu} = R_{\mu\nu} - \frac{1}{2}Rg_{\mu\nu}$ is the Einstein's tensor. In co-moving co-ordinate system, $v_\mu v^\mu = -1$, and the corresponding field equations are

$$\frac{\dot{a}^2}{a^2} + \frac{k}{a^2} = \frac{8\pi G}{3}\rho; \quad 2\frac{\ddot{a}}{a} + \frac{\dot{a}^2 + k}{a^2} = -8\pi G p. \tag{3}$$

Note that the continuity equation of the fluid, which is obtained from the so-called Bianchi identity, $T^{\mu\nu}{}_{;\mu} = 0$, is expressed as $\dot{\rho} + 3\frac{\dot{a}}{a}(\rho + p) = 0$, is not an independent equation, rather, may be found from suitable combination of the field equations. Combination of the field equations (3), also reads

$$\frac{\ddot{a}}{a} = -\frac{4\pi G}{3}(\rho + 3p). \tag{4}$$

Since there are three unknowns, viz., $a(t), \rho(t)$ and $p(t)$ out of two independent equations, a third relation is necessary. For instance, if the energy density is dominated by one component fluid, it is provided by an equation of state of the fluid, viz., $p = \omega\rho$, where $\omega$ is the equation of state (EOS) parameter. In a given state, $\omega$ is constant, (e.g., $\omega = 0$ for dust and $\omega = \frac{1}{3}$ for radiation), but in general it is not. The general solutions, in view of constant value of state parameters were presented by Friedmann (Friedmann 1922, 1924) and later independently by Lemaître (Lemaître 1927) in the form, $a = a_0 t^{\frac{2}{3(1+w)}}$. Therefore, in the pure radiation era, the scale factor behaves as, $a = a_0\sqrt{t}$, while in the pure dust era, it behaves like $a = a_0 t^{\frac{2}{3}}$. Now, it is clear from equation (4) that if strong energy condition $(\rho + 3p) \geq 0$ i.e. $\omega \geq \frac{1}{3}$ is satisfied, then the universe will always be in the decelerating phase of expansion. Further, extrapolation of the expansion of the universe back in time leads to infinite density and temperature at a finite time in the past, together with the geometry of space-time, particularly the Kretschmann scalar $(R_{abcd}R^{abcd})$ (Hawking and Ellis 1973). This initial singularity is commonly known as "Big Bang", which is a singularity extending through all space at a single instant of time. This is the well-known FLRW model, or more commonly known as 'the standard model of cosmology'.

The success of the standard model of cosmology are the following. Firstly, Friedmann's solution of Einstein's equation implies an expanding universe (Friedmann 1922, 1924). Finding luminosity distance $d_L$ versus redshift $z$ relationship of galaxies, expansion of the universe was experimentally confirmed by Hubble in 1929 (Hubble 1929). Secondly, it predicts the existence of `Cosmic Microwave Background Radiation' (CMBR). It is supposed that nearly three minutes after the big bang, nucleosynthesis was initiated, i.e., hydrogen and helium nuclei were formed from protons and neutrons. Approximately, for the next $10^5$ years, the universe was in a radiation dominated phase, in which matter and radiation were strongly coupled through Thompson scattering (Dodelson 2003). With the expansion, the universe cools and at around 3000 K, protons and electrons combine to form neutral hydrogen. At this era, known as recombination era , photons were decoupled and free stream. Thus, the photons decoupled from matter at the last scattering surface and could travel almost unhindered till the present day, loosing energy continuously as the universe expands. This radiation, known as CMBR, was first observed by Penzias and Wilson (Penzias and Wilson 1965) in the microwave region, and is having a temperature of 2.725 K. Thirdly, the abundance of the light atomic nuclei observed in the present universe agrees fairly well with the predictions of standard model. The prediction gives the correct abundance ratio of $\frac{He^4}{H} \sim 0.25$, $\frac{D^2}{H} \sim 10^{-3}$, $\frac{He^3}{H} \sim 10^{-4}$ and $\frac{Li^4}{H} \sim 10^{-9}$ (by mass, and not by number).

Despite tremendous success, standard model of cosmology suffers from pathology. Firstly, the problem of initial singularity as already discussed, has been found to be incurable till date. Next, large entropy per baryon ($\sim 10^8$) has been observed in the present day universe. The reversible adiabatic process, assumed in the standard model of cosmology has no answer to this problem.

As mentioned, thermodynamics of the standard model has been developed under the assumption of reversibility and adiabaticity. The entropy density $s$ can be derived via the second law of thermodynamics, $dE = TdS - PdV$, where $V \propto a^3$ ($a$ being the scale factor) is a co-moving volume with energy $E = \rho V$ and entropy $S = sV$. This can be expressed as

$$d\rho = (sT - \rho - P)\frac{dV}{V} + Tds \qquad (5)$$

And, because $\rho$ depends on $T$, equation (5) implies that, $s = \frac{\rho+P}{T}$. So, for reversible adiabatic universe, total entropy remains constant.

## 3   Dissipative mechanism

In the early eighties it was realized that the standard model cosmology has no answer to the observed large entropy per baryon ($\sim 10^8$) in the present day universe. The inclusion of dissipative terms, such as those due to viscosity and heat flow in the constituent fluid of a cosmological model, introduces several interesting features in its dynamics. As pointed by Misner (Misner 1967), there are at least two stages when irreversible processes certainly cannot be ignored, viz., when the neutrinos decouple and when photons decouple. At these stages matter behaves as a viscous fluid and this might play an important role in galaxy formation as pointed out by Ellis (Ellis 1971). The importance of the dissipative effects increased more since, it could provide possible explanation for the very large entropy per baryon, being observed in the present universe. As already mentioned, this observed fact could not be explained by means of perfect fluid cosmologies. In the homogeneous and isotropic Robertson-Walker (RW) model, the matter is shear-free and non-conducting, so that the dissipation can arise only through the mechanism of bulk viscosity. Weinberg (Wienberg 1971), Treciokas and Ellis (Treciokas and Ellis 1971), and Nightingale (Nightingale 1973) considered the problem of entropy generation due to bulk viscosity in RW model. The entropy estimated in all these calculations indicate that this particular bulk viscosity coefficient $\zeta$, could not be solely responsible for the entropy associated with the background radiation. The other effect can be understood from the consideration of Einstein's field equations. The general energy-momentum-stress tensor for a viscous fluid (see e.g. (Landau and Lifshitz 1959; Wienberg 1972)) is given by

$$T_{\mu\nu} = (\rho + \bar{p})v_\mu v_\nu + \bar{p}g_{\mu\nu} - \eta U_{\mu\nu} \tag{6}$$

where, $U_{\mu\nu} = v_{\mu;\nu} + v_{\nu;\mu} + v_\mu v^\alpha v_{\mu;\alpha} + v_\nu v^\alpha v_{\mu;\alpha}$ (7)

and, $\bar{p} = p - \left(\zeta - \frac{2}{3}\eta\right)\Theta$ (8)

In the above, $\bar{p}$ is the effective pressure, $\eta$ is the coefficient of shear viscosity, and $\Theta$ is the expansion scalar. Treciokas and Ellis (Treciokas and Ellis 1971) showed that the scale factor $a$ of the Robertson-Walker metric, in the presence of bulk viscosity takes the form, $a^{\frac{3\gamma}{2}} = \frac{2}{3\zeta}\left(e^{\frac{3\zeta t}{2}} - 1\right)$, where $\gamma$ is defined by their equation of state, $p = (\gamma - 1)\rho$, and thus $1 \leq \gamma \leq \frac{4}{3}$.

In homogeneous and isotropic cosmological model with non-vanishing bulk viscosity, the nature of singularity has been discussed by Heller, Klimek and Suszycki (Helleret al. 1973), Heller and Suszycki (Heller and Suszycki 1974), and Murphy (Murphy 1973). While in the first two works, the bulk viscosity was assumed to be a constant, Murphy, choosing flat space a-priori, ($k = 0$), assumed it to be proportional to the matter density $\rho$ (i.e. $\zeta \propto \rho$). Singularity free solutions were obtained in such models in the sense that it occurs only at infinite past. Later Banerjee and Santos (Banerjee and Santos 1984) extended Murphy's work for a more general relation $\zeta \propto \rho^n$ and, $k = 0, \pm 1$, which again resulted in singularity free models, in some special cases. However such singularity free models were obtained paying an unphysical price, violating the Hawking Penrose energy condition (Hawking and Ellis 1973).

In the meantime, Cadarni and Fabri (Cadarni and Fabri 1978) argued that the causal structure of RW metric does not allow any kind of communications in faraway parts of the universe. So isotropy although valid at present, looks rather artificial in the early universe. In this context, Belinskii and Khalatnikov (Belinskii and Khalatnikov 1975) considered cosmological solution of the Einstein's equations for the anisotropic Bianchi-I model, where the shear and bulk viscosity coefficients were assumed to be power functions of the energy density, $\rho$. They studied the asymptotic behaviour of the solutions for the final stage of the collapse. One of the important observations was that, viscosity is unable to remove the cosmological singularity, although it introduces qualitatively new elements into the character of the singularity. In the expanding model, the influence of matter near the initial singularity is found to be very small and the metric is determined by the free space Einstein's equation, unlike that in the perfect fluid cosmology, where we encounter infinite energy density at the beginning. The matter density vanishes at the initial instant and subsequently in the course of expansion, it grows. The idea was modified by Banerjee and Santos (Banerjee and Santos 1983) to obtain exact solutions for viscous fluid in Bianchi I and II models with the simplifying assumption that the ratio of shear to expansion $\left(\frac{\sigma}{\Theta}\right)$ has a fixed magnitude throughout the evolution. The results are, however, quite different from those of Belinskii and Khalatnikov (Belinskii and Khalatnikov 1975). Banerjee and Santos (Banerjee and Santos 1975) also discussed the entropy change due to dissipative processes. If $\epsilon$ be the entropy change per baryon, $s$ be entropy per unit volume and $n$ be the number density per baryon, then we have the following relations

$$\dot{\epsilon} = -\frac{1}{nT}[\dot{\rho} - (\rho+p)\Theta]; \quad \frac{\dot{s}}{s} = \frac{\dot{\rho}}{\rho+p}, \text{ and}, \frac{\dot{S}}{S} = \frac{\dot{\rho}}{\rho+p}\Theta = \frac{4\eta\sigma^2 + \zeta\Theta^2}{\rho+p} \quad (9)$$

where, $S = a^3 s$ is the total entropy. However, for perfect fluid, $\eta = \zeta = 0$, so that $\dot{S} = 0$. So, the total entropy remains conserved, as argued before. The above expression gets modified in the presence of heat flux.

Later, Banerjee, Duttachowdhury and Sanyal studied the effect of both bulk and shear viscosities in Bianchi-I and Bianchi-II cosmological models (Banerjee et al. 1985, 1987). In these models it was shown in general, how the dynamical importance of the shear and the fluid density given by $\frac{\sigma}{\Theta}$ and $\frac{\rho}{\Theta}$ respectively change in the course of evolution. The relation which shows the time rate of variation of these two quantities is given by

$$\frac{d}{dt}\left(\frac{\sigma^2}{\Theta^2}\right) = -\frac{d}{dt}\left(\frac{\rho}{\Theta^2}\right) = \left(\frac{\sigma}{\Theta}\right)^2 [3(\rho-p)\Theta^{-1} + 3\zeta + 4\eta] \quad (10)$$

So, in an expanding model, for reasonable physical requirements viz. $\rho \geq p > 0$, $\zeta > 0$, $\eta > 0$, one finds, $\frac{d}{dt}\left(\frac{\rho}{\Theta^2}\right) < 0$, while, $\frac{d}{dt}\left(\frac{\sigma^2}{\Theta^2}\right) > 0$. The shear propagation relation is given by $(\sigma^2)^{\cdot} = -2(2\eta + \Theta)\sigma^2$, which shows that $\sigma^2$ decreases with time in the process of expansion. Further bulk viscosity has also been found to play a significant role in the process of shear dissipation mechanism. However, nothing much definite could be said for a collapsing model. The total entropy was found to increase throughout the evolution for nonnegative values of matter density. This is of course consistent with the anomalously high entropy per baryon observed in the present day universe. Moreover from Raychaudhury equation, it becomes evident that with energy conditions being satisfied, there cannot be any occurrences of bounce from a minimum volume, and so one cannot avoid the initial singularity. It should be noted at this point that the singularity free solution due to Murphy (Murphy 1973), however, violates the above mentioned energy condition. The explicit form of the total entropy obtained in a special case where $\zeta$ and $\eta$ are assumed to be constant quantities is, $S = S_0 \left(\frac{1}{3}e^{3\zeta_0 t} - A^2 e^{-4\eta_0 t}\right)^{\frac{1}{2}}$, and so at a finite instant the entropy is zero, but the total entropy, however, increases indefinitely as $t \to \infty$. So, in this model, the entropy is found to always increase and the singularity of zero volume is unavoidable. The model isotropizes in the asymptotic limit and the Hawking-Penrose energy condition (Hawking and Ellis 1973) is found to be satisfied.

The effect of axial magnetic field in the presence of both the viscous co-efficient were also studied by Banerjee and Sanyal (Banerjee and Sanyal 1986) and Ribeiro and Sanyal (Ribeiro and Sanyal 1987) in Bianchi I, Bianchi III and Kantowski-Sachs models and in Bianchi $VI_0$ model respectively. The magnetic field is represented by the only non-vanishing component of the electromagnetic field tensor, $F_{23} = A\psi(\theta)$, where $\psi(\theta) = \sin\theta, 1$ or $\sinh\theta$ according as the metric is of the Kantowski-Sachs, Bianchi I and Bianchi III type respectively. It was found that the existence of the magnetic field does not alter the fundamental character of the initial singularity.

Another aspect of the dissipative process is the heat flux, which is often discussed in the context of cosmology (see e.g. (Wienberg 1972)). If the present entropy of the universe is not due to bulk viscosity then possibly it is produced by the effects of shear viscosity and/or heat conduction. The nonrelativistic treatment of this process results in the inclusion of a suitable term in the energy momentum tensor, which can again be extended to a generally co-variant form. The procedure is to introduce appropriate dissipative terms in the energy momentum tensor and study Einstein's field equations along with the exact solutions. The most general energy momentum tensor therefore reads

$$T_{\mu\nu} = (\rho + \bar{p})v_\mu v_\nu + \bar{p}g_{\mu\nu} - \eta U_{\mu\nu} + q_\mu v_\nu + q_\nu v_\mu \tag{11}$$

The study of Bianchi V cosmological model with heat flux by Banerjee and Sanyal (Banerjee and Sanyal 1988), couldn't modify earlier results.

In the early eighties, some additional problems of standard model cosmology, by the name of "flatness", "horizon" and "structure formation" came into picture, which can't be solved simply from dissipative phenomena.

## 4 Flatness, horizon and structure formation problem: need for Inflation

As mentioned, the standard model has been found to suffer from some additional problems viz., flatness problem, horizon problem, structure formation problem etc. in the early universe. In this section, for the sake of completeness, let us qualitatively discuss these problems together with their resolution.

The flatness problem is analogous to cosmological fine-tuning problem of the universe. It arises from the observation that some of the initial conditions (e.g., the density of matter and energy) of the universe require to be fine-tuned to very 'special' values, and that a small deviation from these values would have had massive effects on the nature of the universe at the present time. Friedmann's equation (3) may be expressed in terms of density parameter, $\Omega = \frac{\rho}{\rho_c}$, ( $\rho_c = \frac{3H^2}{8\pi G}$ being the critical density at any instant, required to produce a flat universe), as

$$\Omega - 1 = \frac{k}{a^2 H^2}, \quad \text{so that,} \quad |\Omega - 1| \propto \begin{cases} \sqrt{t} & \text{Radiation era} \\ t^{\frac{2}{3}} & \text{Matter era} \end{cases} \quad (12)$$

In the above, $H = \frac{\dot{a}}{a}$, is the Hubble parameter. Although, $\Omega = 1$ is an unstable critical point, it was evident at the early times, that it should be close to one, since $\Omega_B$ was measured to be of the order of 1. Recent observations of course unambiguously constrain the present value of the density parameter very close to 1 (Spergel 2007). Even slightest deviation in the value of $\Omega$ from 1 leads to a huge difference from 1 at the very early universe. Particularly, $\Omega = 1$, at present time ($t \approx 14$ Gyr), implies that it must have been incredibly close to one at early times. In view of standard model, $\Omega_{t=100s} = 1 \pm 10^{-11}$, and $\Omega_{Planck} = 1 \pm 10^{-62}$. This leads cosmologists to question how the initial density came to be so closely fine-tuned to this very 'special' value.

The Cosmic microwave background radiation (CMBR) supposed to have been emitted from the last scattering surface, is spatially homogeneous and isotropic. This homogeneity indicates causal connection among every point on the last scattering surface. But standard cosmology does not find any clue to these causal connections. This is because, light signal travels a finite distance in a finite time and casually connect points within this finite distance which is called the horizon. In view of the standard model, the sky splits into $1.4 \times 10^4$ patches, which were never casually connected before emitting CMBR. So the puzzle, why the universe appears to be uniform beyond the horizon, is called horizon problem.

The universe, as is now known from observations of the cosmic microwave background radiation, began from a state of hot, dense, nearly uniform distribution, approximately 13.8 billion years ago (Spergel 2007). However, looking at the sky today, we see structures on all scales, from stars and planets to galaxies and, on much larger scales, cluster of galaxies, and also enormous voids between galaxies. It would have been generated from some seed of density perturbation, in the early universe. But the standard model does not admit such perturbations.

There are also some associated problems in regard of unwanted relics, viz. monopoles, topological defects like domain walls, cosmic strings, gravitino, the spin $-\frac{3}{2}$ partner of graviton, moduli – the spin 0 particles etc. However, all these stem from our present understanding of particle physics. So we leave the discussion, stating that, the resolution to the main three problems resolves these issues also.

## 4.1 Inflationary scenario

All the problems of the standard model discussed above have been alleviated invoking one powerful concept: inflation. Alan Guth (Guth 1981) proposed an intermediate phase of expansion of the universe, the so called inflationary phase ($\ddot{a} > 0$, with, $\rho_\phi + 3p_\phi < 0$, as the scalar field, inflaton $\phi$ drives inflation), at GUT epoch ($10^{-36}$ sec), before the standard Big Bang of the very early universe. In the inflationary phase the universe underwent a rapid exponential expansion, for a very short period of time, attributed by a scalar field which contributes to the energy-momentum tensor of the Einstein's field equation. This is known as inflationary model of the universe.

Inflation works in a fairly simple manner. If the universe expands with a scale factor, $a \propto e^{Ht}$ (for, $\rho_\phi + p_\phi = 0$), that grows more rapidly than the velocity of light, a very small region, initially in thermal equilibrium, can easily grow to encompass our entire visible universe at last scattering. The mutual thermalization of the 'apparent decoupled regions' is then obvious, and there is no need to assume initial homogeneity. Two points which were casually connected in the very early expanding universe have fallen large apart so their past light cones never intersect even if they are extended back to the last scattering surface. Thus the horizon problem can be solved.

Inflation removes the pathology due to the flatness problem by blowing up the scale factor to huge proportions, like inflating a balloon to a larger volume smears the wrinkles and flattens the surface. An accelerating scale factor drives the density parameter $\Omega$ towards 1, without any fine-tuning of initial conditions. Thus the flatness problem is solved by the fact that the accelerated expansion 'blows away' the curvature.

If the universe were inflated right after the GUT-era, monopoles, together with all other unwanted thermal relics are simply blown away by the dilution of the energy density caused by inflation, and are untraceable at present.

Inflation is believed to be caused by a self-interacting quantum field, exhibiting vacuum fluctuations. Since, the length scales of the fluctuations leave the Hubble scale during inflation, they freeze to become classical. At Hubble scale re-entry (after inflation stops), these fluctuations form the seeds of perturbation, which is responsible for structure formation. Regions with a higher density accumulate matter and regions with a lower density lose some of their energy content to the higher density regions according to the Jeans-mechanism. A review on these issues may be found in (Sachs and Wolfe 1967; Bardeen 1980; Mukhanov et al. 1992). In figure-1, the observed seeds of perturbation in the form of temperature fluctuation by Planck mission are clearly visible.

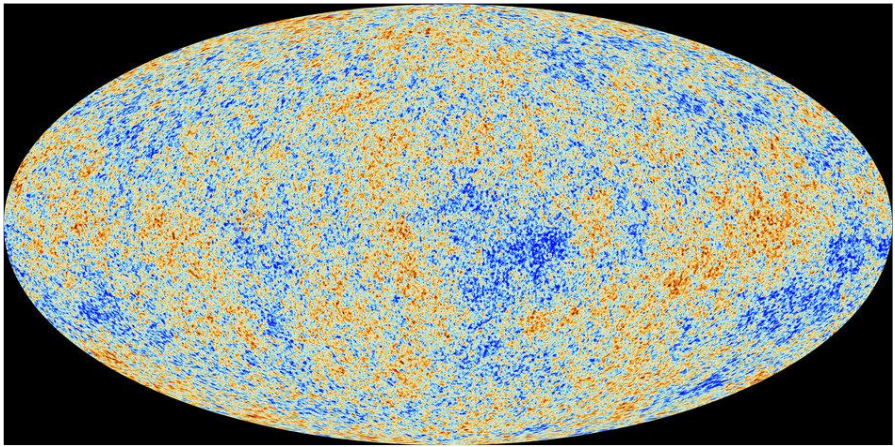

**Figure.1 Temperature fluctuation of CMB at last scattering surface, received from Planck.**

Despite its success, Guth's old inflationary model suffers from a problem of its own second order phase transition, called the graceful exit problem. The success of the standard model in explaining BBN and CMBR suggests that the universe must have started from a state of very hot dense plasma state. Inflation is a period of super cooled expansion, and the temperature drops by a factor of $10^5$ or so, to nearly 1000 K. This temperature doesn't allow nucleosynthesis. As a result, CMBR might not have been present and structures would not have been formed. This indicates that after the graceful exit from inflation by some means, the temperature of the universe must have increased (reheating) to give way to the standard Big-Bang, which is now by no means a singularity, but a hot thick soup of plasma. However, Guth's inflation never ends.

This problem was addressed in a number of models such as the new inflationary model (Linde 1982; Albrecht and Steinhardt 1982), chaotic inflationary model (Linde 1983), extended inflationary model (La and Steinhardt 1989; Mathiazhagan and Johri 1984; Sanyal and Modak 1992; Barrow 1995), hyperextended inflationary model (Steinhardt and Accetta 1990), and in Starobinski's model of curvature induced inflation - without phase transition (Starobinsky 1980). However, all these models suffer from some sort of demerits. The new inflationary and chaotic inflationary models require fine tuning of the effective potential parameter. This problem was removed in the extended inflationary model where the first order phase transition yields variation in the gravitational constant as in Jordan-Brans-Dicke theories (Dicke 1962; Brans and Dicke 1961; Dicke 1962; Bergmann 1968; Wagner 1970). The extended inflationary model on the other hand, finds its problem in setting a very small value of the Brans-Dicke parameter $\omega < 50$ so that the distortion in CMBR is negligible. This is against the present observed value of $\omega > 40000$ (Bertotti et al. 2003). However, some inflationary models with noniminimally coupled scalar tensor theory (Kalloh, Linde and Roest. 2014), (Bezrukov and Shaposhnikov 2008) appear to be free from pathology. Nevertheless, slow-roll approximation is achieved in these models following scalar-tensor equivalence to Einstein's frame, while physical equivalence between the two frames is debatable.

For detailed discussion on different models of inflation, calculation of power spectrum, reheating, observational constraints and other issues, we refer to the excellent review (Bassett et al. 2006). However, particle creation at the expense of gravitational field is also a strong candidate of inflationary scenario. We therefore keep our focus on the particle creation phenomena in the following section.

## 5 Thermodynamics of irreversible particle creation phenomena

In the absence of a viable quantum theory of gravity, two directions were pursued to understand the behaviour at or near Planck's era. One is through quantization of the cosmological equation, viz. the Hamilton constraint equation, known as Wheeler-deWitt equation and the momentum constraint equations. These constraints are the outcome of reparametrization invariance (diffeomorphic invariance, to be more specific) of the theory of gravitation. The other is to explore the era, when all the fields but gravity are quantized. This second option is known as quantum field theory in curved space time (QFT in CST). The missing link between the very early and late stage of cosmological evolution has been partly illuminated following the study of QFT in CST. In this framework, quantum field theory in space-times is described by classical metrics, as in general relativity, in a regime where both theories are valid. The equation then takes the form, $G_{\mu\nu} = \kappa < T_{\mu\nu} >$, where, $< T_{\mu\nu} >$ is the expectation value of the energy momentum tensor comprising all possible fields. The epoch-making event in the study of QFT in CST, is the Hawking's invention of black hole radiation, and relating the entropy of the black hole with the area of the horizon (Hawking1974, 1975). Another remarkable physical outcome of QFT in CST is the phenomenon of gravitationally-induced spontaneous creation of quanta in curved space-times, in the cosmological context of an expanding universe. Such spontaneous creation of particles at the expense of gravitational fields were extensively explored during the last century by Parker and his collaborators (Parker 1968, 1969, 1971; Papastamatiou and Parker 1979). For further details, we refer to the two famous books by Birrell and Davies (Birrell and Davies 1982) and Mukhanov and Winitzki (Mukhanov and Winitzki 2007). However, we mention at this stage that the tensor contributions to the quadrupole in the cosmic microwave background imply that the energy density of the inflaton field satisfies, $V_{60} = 6 \times 10^{-11} M_p^4$, where $M_p$ is the Plank's mass. Thus, inflation is treated as a low energy, classical phenomenon. Hence, gravity was classical during inflation.

Cosmological consequence of the particle-creation mechanism is studied taking into account an explicit phenomenological balance law for the particle number (Prigogine et al. 1989; Prigogine 1989; Calvao et al. 1992; Lima et al. 1996; Zimdahl and Pavón 1993; Zimdahl et al. 1996) in addition to the familiar Einstein's equations. In view of such a balance law, Prigogine et al. (Prigogine et al. 1989; Prigogine 1989) successfully explained the cosmological evolution of the early universe.

Let us formulate the balance equation in connection with particle-creation phenomena. Adiabatic cosmological evolution in the presence of particle creation can be treated in the open system, and so the first law of thermodynamics gets modified to

$$d(\rho V) + p_m dV - \frac{h}{n} d(nV) = 0 \tag{13}$$

where, $\rho$, $p_m$, $V$, $n$ and $h$ are the total energy density, the true thermo dynamical pressure, any arbitrary co-moving volume, the number of particles per unit volume and the enthalpy per unit volume respectively. Here, the system receives heat only due to the transfer of energy from gravitation to matter. So, creation of particles acts as a source of internal energy. Thus, for adiabatic transformation, the second law of thermodynamics reads

$$TdS = d(\rho V) + p_m dV - \mu d(nV) \tag{14}$$

Combining the above two equations (13) and (14), we have

$$TdS = \frac{h}{n}d(nV) - \mu d(nV) = T\epsilon dN \tag{15}$$

To derive the above expression, we have used the usual expression for the chemical potential as, $\mu n = h - Ts$. Here, $s$ and $\epsilon$ stand for the entropy per unit volume and specific entropy. Thus, we observe that the second law of thermodynamics, viz., $dS \geq 0$, requires $dN \geq 0$, and the reverse process is thermodynamically impossible, i.e., particle can only be created and cannot be destroyed. Further, expressing $S$ in terms of $\epsilon$, the above equation can also be expressed as $TNd\epsilon = 0 \Rightarrow \dot{\epsilon} = 0$.

Hence, in the adiabatic particle-creation phenomena, entropy increases, while the specific entropy remains constant. The first law given by equation (13) can also be expressed as, $Vd\rho + \rho dV + p_m dV - hdV - \frac{hV}{n}dn = 0$, which reduces to

$$Vd\rho - \frac{hV}{n}dn = 0 \Rightarrow \dot{\rho} = h\frac{\dot{n}}{n} \tag{16}$$

Now, the energy–momentum tensor $T^{\mu\nu}$ taking into account the conservation law incorporating creation phenomena reads,

$$T^{\mu\nu} = (\rho + p_m + p_{cm})u^\mu u^\nu - (p_m + p_{cm})g^{\mu\nu}, \qquad T^{\mu}_{\nu;\mu} = 0 \tag{17}$$

where, $\rho = \rho_m + \rho_{cm}$ is the total energy density, $p = p_m + p_{cm}$, $p_{cm}$ being the creation pressure and $u^\mu$ is the component of the four-velocity vector. The above energy conservation law in homogeneous cosmological models may be expressed as,

$$\dot{\rho} + \Theta(\rho + p_m + p_{cm}) = 0 \tag{18}$$

where, $\Theta$ is the expansion scalar. Plugging in the expression for $\dot{\rho}$ from (16), in the above equation, one obtains

$$p_{cm} = -\frac{\rho + p_{cm}}{\Theta}\left(\Theta + \frac{\dot{n}}{n}\right) = \frac{\rho + p_{cm}}{\Theta}\Gamma \tag{19}$$

where, $\Gamma = \Theta + \frac{\dot{n}}{n}$ is the creation rate. Now, the inequality

$$dN = d(nV) \geq 0 \Rightarrow \dot{n} + 3Hn \geq 0 \tag{20}$$

is compatible with $H \geq 0$, $H = 0$ and $H \leq 0$. However, in the case of a de Sitter universe, in which $\dot{\rho} = 0$, the above relation reduces to $H \geq 0$ by virtue of the relation $\dot{\rho} = \frac{\dot{n}}{n}(\rho + p)$. So, only an expanding de Sitter universe is thermodynamically possible.

### 5.1 Inflation and entropy burst

Now, in order to explore non-traditional cosmology which includes particle creation, Prigogine et al. (Prigogine et al. 1989; Prigogine 1989) presented a simple phenomenological model. This model provides a cosmological history which evolves in three stages: first, a creation period which drives the cosmological system from an initial fluctuation of the vacuum to a de Sitter space, which is the second stage of cosmic evolution. This de Sitter space exists for the decay time $\tau_d$ of its constituents. Finally, a phase transition turns this de Sitter space into a usual Robertson-Walker (RW) universe, which extends to the present. At that time, due to lack of knowledge of the presently accelerated universe (this will be discussed in the following section), entropy creation was thought to occur only during the two first cosmological stages, while the RW universe was assumed to evolve adiabatically on the cosmological scale. Prigogine et al. (Prigogine et al. 1989; Prigogine 1989) expressed the irreversible creation phenomena in terms of the Hubble function $H$ as follows:

$$\frac{1}{a^3}\frac{d(na^3)}{dt} = \alpha H^2 \geq 0, \quad \text{with,} \quad \alpha \geq 0 \tag{21}$$

Now in view of the simple additional relation, $\rho = Mn$, the pressure vanishes, $p = 0$. Hence, for $\alpha = 0$, one can recover the usual RW description with its typical big-bang singularity, as the solution for the spatially flat Einstein's equation $\kappa\rho = 3H^2$ ($\kappa = 8\pi G$). However for $\alpha \neq 0$, one obtains

$$p = 0, \quad \rho = \frac{3H^2}{\kappa}, \quad \text{and,} \quad \frac{1}{na^3}\frac{d(na^3)}{dt} = \frac{\alpha\kappa M}{3} \geq 0 \tag{22}$$

This leads to $N = N_0 e^{\frac{\alpha\kappa Mt}{3}}$, and, $a(t) = \left[1 + C\left(e^{\frac{\alpha\kappa Mt}{6}} - 1\right)\right]^{\frac{2}{3}}$ (23)

where, $C = \frac{9}{\kappa M\alpha}\sqrt{\frac{\kappa Mn_0}{3}}$. Thus, the universe emerges without singularity, ($a \neq 0$, at $t = 0$, and nothing blows) with a particle density $n_0$ describing the initial Minkowskian fluctuation. It therefore follows that the presence of dissipative particle creation ($\alpha \neq 0$) leads to the disappearance of the big-bang singularity. In other words, this singularity is structurally unstable with respect to irreversible particle creation. Hence, such a cosmological model starts from instability ($n_0 \neq 0$) and not from a singularity.

After a characteristic time, $\tau_c = \frac{6}{\alpha\kappa M}$ the universe reaches a de Sitter regime characterized by, $a_d(t) = C^{\frac{2}{3}}e^{\frac{2t}{3\tau_c}}$, $H_d = \frac{2}{3\tau_c}$, and $n_d = \frac{\kappa M}{27}\alpha^2$. The de Sitter stage survives during the decay time $\tau_d$ of its constituents and then connects continuously to a usual RW universe characterized by a matter-energy density $\rho_b$ and radiation energy density $\rho_\gamma$, related to the scale factor by, $\kappa\rho_b = \frac{3A}{a^3}$, $\kappa\rho_\gamma = \frac{3B}{a^4}$ and hence, $\rho_\gamma = \frac{\pi^2}{15}T^4$. Here, $A$ and $B$ are constants related to the total numbers $N_b$ of baryons and $N_\gamma$ of photons in a volume, $a^3$, and T is the blackbody radiation temperature. The connection at the decay time $\tau_d$ between the de Sitter and the matter-radiation regimes fixes the constants as $A \simeq 2H_d^2 C^2 e^{2H_d\tau_d}$ and, $B \simeq H_d^2 C^{\frac{8}{3}} e^{4H_d\tau_d}$. This implies that the specific entropy S per baryon is a constant, given by,

$$S = \frac{n_\gamma}{n_b} = \frac{\zeta(3)}{3\pi^2}\left(\frac{45}{\pi^2}\right)^{\frac{3}{4}} \kappa^{\frac{1}{4}} m_b \left(\frac{3\tau_d}{2}\right)^{\frac{1}{2}} e^{\frac{2\tau_d}{3\tau_c}} \tag{24}$$

where, $m_b$ stands for the baryonic mass. Both the quantities $\tau_d$ and $\tau_c$ can also be expressed in terms of one single parameter (Gunzig et al. 1987), viz., the mass $M$ of the produced particles. These values are, $\tau_d \simeq 2.5 \left(\frac{M}{M_p}\right)^3 \tau_p$ and $\tau_c \simeq 1.42 \left(\frac{M}{M_p}\right)^2 \tau_p$, where $M_p$ and $\tau_p$ are the Planck mass and the Planck time, respectively. Further, the correct observed values of S, can be obtained for values of the mass M very close to the one found in view of quantum field theory in curved space time (QFT in CST), which is $M = 53.3\, M_p$ (Spindel 1981). For example, even in the presently considered over simplified model, taking, $\frac{M}{M_p} = 49.5$, the de-Sitter phase is reached within $\tau_c \approx 10^{-39}$s., and within a very short period of time, $\tau_d \approx 10^{-38}$s., for which the de-Sitter phase lasts, $S \approx 10^8$, i.e. there is a burst of entropy. Additionally, the present black body temperature may also be found from the continuity requirements as

$$T_0(°K) \simeq 2.82 \times 10^{-9}\left(\frac{H_0}{75\text{km.s}^{-1}.\text{Mpc}^{-1}}\right)^{\frac{2}{3}} \left(\frac{M}{M_p}\right)^{\frac{1}{3}} e^{0.3926\frac{M}{M_p}} \tag{25}$$

where, $H_0$ is the presently observed value for the Hubble parameter: $H_0 = 73.8 \pm 2.5\text{ km.s}^{-1}.\text{Mpc}^{-1}$ (Reiss et al.2011). Taking, $H_0 = 72.2\text{ km.s}^{-1}.\text{Mpc}^{-1}$, the observed black body radiation temperature can also be obtained with the same above ratio of masses:

$$\frac{M}{M_p} = 49.5, \quad \text{yields,} \quad T_0 = 2.7255°K \tag{26}$$

At this end, we observe that the energy transfer from space-time curvature to matter, is an irreversible process leading to a burst of entropy associated with the creation of matter. It follows therefore that the distinction between space-time and matter is provided by entropy creation. As already mentioned in the case of the de Sitter universe, only expansion is thermodynamically possible. The universe always develops through a de Sitter stage. As a result, there is indeed a direct relation between the existence of cosmological entropy and the expansion of the universe. Later, Zimdahl, Triginer and Pavón (Zimdahl and Pavón 1993; Zimdahl et al. 1996) studied exponential and power law inflationary scenarios, under creation phenomena. The computation of the spectrum of scalar and tensor perturbation under slow-roll inflation has been exhaustively investigated by Parker (Agullo and Parker 2011). The observations (Komatsu et al 2011, Planck collaboration 2014, and 2016) of the scalar to tensor ratio, $r = \frac{P_T}{P_R} < 0.14$, where, $P_T$ and $P_R$ are the tensor and the curvature perturbations respectively, and the spectral index $0.96 < n_s < 0.984$, only constrain the average number of initial quanta. On the contrary, the observed little non-gaussianities in the distribution of the perturbations (Planck collaboration 2014, Planck collaboration 2016), is interpreted as a consequence of a non-vacuum initial state.

# 6 A sharp turn in Cosmology: Type Ia Supernovae observation and late-time acceleration

Type Ia supernovae (SNeIa) are ideal astronomical objects for distance determination. These objects are created through the explosion of accreting white dwarf stars when they reach the Chandrasekhar limit. Since they all have the same progenitors and are triggered through a consistent underlying mechanism, so they are assumed to have constant absolute magnitude and hence may be treated as standard candles. Indeed it was the observations of SNeIa that confirmed the acceleration of the universe by two independent teams of Riess et al. (Riess et al. 1998) and Perlmutter et al. (Perlmutter et al. 1999).

The analysis of SNeIa data are made by plotting its observed distance modulus $\mu (= m - M)$ against redshift $z$ in Hubble diagram and comparing this light curve with the curve obtained from theoretically predicted values. Here, $m$ and $M$ are the apparent and absolute magnitude of luminosity. A physical quantity called luminosity distance $d_L$ is defined as, $d_L = \sqrt{\frac{L}{4\pi F}}$, where $L$ and $F$ are the apparent and the absolute luminosities respectively. Finally, one obtains the relation, $\mu = 5\,Log_{10}\left(\frac{d_L}{Mpc}\right) + 25$. Now to obtain best fit with the experimental luminosity-distance versus redshift curve, vacuum energy is invoked. The Hubble parameter, for the purpose, is expressed in the following convenient form

$$\left(\frac{H}{H_0}\right)^2 = \sum_i \Omega_i^0 (1+z)^{3(1+\omega_i)}$$

$$= \Omega_r^0 (1+z)^4 + \Omega_m^0 (1+z)^3 + \Omega_\Lambda^0 + \Omega_k^0 (1+z)^2 \qquad (27)$$

where, $\Omega_r^0$, $\Omega_m^0$, $\Omega_\Lambda^0$ and $\Omega_k^0$ are the density parameters corresponding to radiation, matter, vacuum energy and curvature at present epoch respectively, $z$ is the redshift parameter and, $\omega_i$ is the corresponding equation of state (EOS) parameter. As already mentioned, $\sum_i \Omega_i^0 = 1$ is strongly supported by observational data. Hence the luminosity distance is given by

$$d_L = \frac{1+z}{H_0} \int_0^z \frac{dz'}{\sum_i \Omega_i^0 (1+z')^{3(1+\omega_i)}} \qquad (28)$$

It has been found that the best fit of the experimental curve with the theoretical one in a two component flat universe, requires vacuum energy density $\Omega_\Lambda^0 = 0.7$ and matter energy density $\Omega_m^0 = 0.3$ (Riess et al. 1998, Perlmutter et al. 1999). This is known as ΛCDM model. Figure 2 shows a plot of distance modulus vs. redshift (Choudhury and Padmanabhan 2005). So the primary conclusion is −universe is filled with (70%) dark energy.

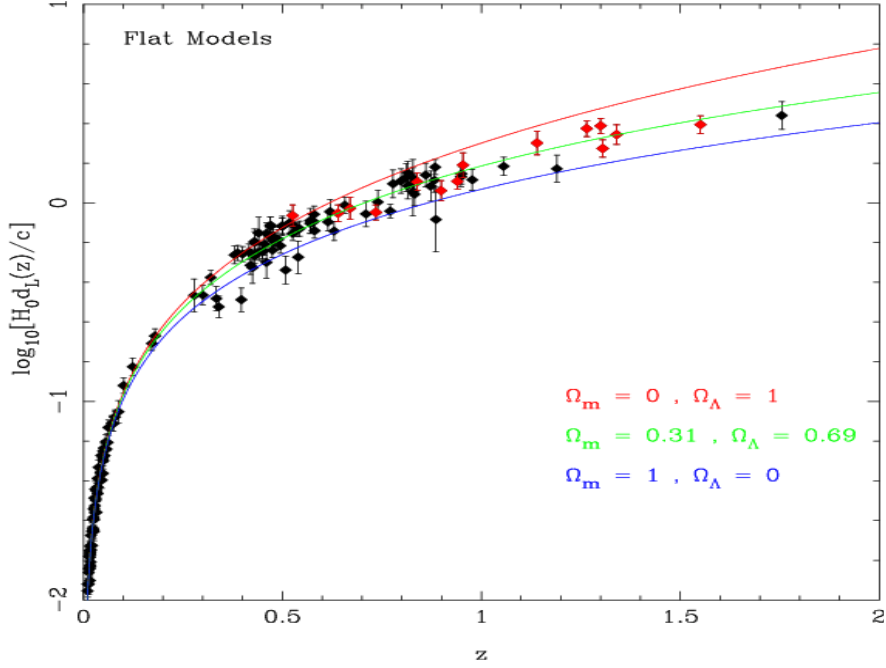

**Figure.2 Comparison between various flat models and the observational data. The observational data points, shown with error-bars, are obtained from the "gold" sample of Riess et al. (Riess et al. 1998). The most recent points, obtained from HST, are shown in red. Courtesy (Choudhury and Padmanabhan 2005).**

Since the observed amount of Baryons present in the universe is only 4%, while SNeIa data reveals that the matter contribution is about 30%, so, naturally there should be about 26% of some exotic kind of matter which interacts only gravitationally with the others. This type of matter is called the "dark matter". Observed peculiar velocities in the galaxies and 21 cm. emission lines of neutral hydrogen instead of stars confirm the presence of dark matter (Peebles 1980, 1993). Further, the seeds of perturbations can produce the presently observed structures by accreting matter in the over-dense regions only in the presence of cold (non-relativistic) dark matter.

## 6.1  Dark energy Models

Einstein's equation must be modified to incorporate recent cosmological observations. Either, one can modify the right hand side of the equation, i.e. the energy-momentum tensor, by incorporating some exotic field – the dark energy or one can also modify the left hand side of the equation by introducing higher order curvature invariant terms, leading to modified theory of gravity. The third option is to consider creation of dark matter at the expense of gravitational field. We briefly discuss below some of these popular models.

**6.1.1 Cosmological constant (ΛCDM) model**

In cosmology, the cosmological constant (Λ) was originally introduced by Albert Einstein to ensure a static universe, which was then a philosophically accepted view. After Hubble's discovery of expanding universe, Einstein abandoned the idea, stating it to be his "greatest blunder". Later, cosmological constant was revived by field theorists as vacuum energy density of the universe. After the discovery of the present acceleration of the universe, in view of collected data from distant supernovae, the cosmic microwave background and large galaxy redshift surveys, it has been confirmed that the mass-energy density of the universe includes around 70% in dark energy. The cosmological constant is the simplest possible candidate for dark energy, since it is constant in both space and time.

In ΛCDM model, the contribution of the dark energy is attributed in the energy-momentum tensor of the Einstein equation so that the general Friedmann equation reads

$$\left(\frac{H}{H_0}\right)^2 = \Omega_r^0(1+z)^4 + \Omega_m^0(1+z)^3 + \Omega_\Lambda^0 \tag{29}$$

for an almost flat present universe, $\Omega_k^0 = 0$. Here, suffix zero stands for the present epoch. It has been found that SNeIa data in Hubble diagram is best fitted with the theoretical curve of ΛCDM model taking, $\Omega_r^0 = 8 \times 10^{-5}$, $\Omega_m^0 = \Omega_B^0 + \Omega_{CDM}^0 = 0.26$ and $\Omega_\Lambda^0 = 0.74$ in the background of FRW metric. But the problem associated with this model is Λ itself. The vacuum energy density, as calculated by the field theorists, is some $10^{120}$ order of magnitude greater than the cosmological constant (Λ) required by the cosmologists to explain late time cosmic acceleration. This is known as cosmological constant problem. Another way of stating the problem is that the observed renormalized cosmological constant is at least 120 orders of magnitude smaller than the quantum corrections, thus requiring an enormous fine-tuning of the bare cosmological constant, once again.

**6.1.2 Scalar-field models**

The cosmological constant corresponds to a fluid with a constant equation of state, $\omega = -1$. However, other models for which the state parameter is dynamical are also supported by different observations. In these models the state parameter evolves and its present value is close to $-1$, or even less. So one can consider a situation in which the equation of state of dark energy changes with time. A dynamical EOS parameter has the advantage of providing a possible solution for the dark energy problem alleviating the coincidence problem and the fine tuning problem. The question why the dark energy possesses such a very small value at present time is called the coincidence problem. Now, if we find a model with an evolution such that the EOS parameter of dark energy becomes dominant at late times independently of the initial conditions, then we have an answer to the coincidence problem. Note that dark energy could not have been dominant in the early universe, because in that case structures like galaxies could not have been formed. Therefore it is convenient to search for models with a tracker behaviour, in which the dark energy density closely tracks the radiation density until very recently. After this epoch the scalar field has to start behaving as dark energy, eventually dominating the universe. Secondly we have a solution for the fine tuning problem, if the field, which creates an EOS parameter, naturally arises from particle physics and gives exactly the energy density equal to the critical energy density at late times. These considerations motivate a search for a dynamical dark energy model caused by some exotic field. So far, a wide variety of scalar-field dark energy models have been proposed, viz., Quintessence model (Ratra and Peebles 1988; Peebles and Ratra 1988; Frieman et al. 1995; Caldwell et al. 1998; Zlatev et al. 1999), K-essence model (Armendáriz-Picón et al. 1999; Garriga and Mukhanov, 1999; Chiba et al. 2000; Armendáriz-Picón et al. 2000, 2001), Phantom model (Caldwell 2002; Sahni and Shtanov 2003; Alam and Sahni 2002; Elizalde et al. 2004; Caldwell et al. 2003), Tachyon field (Sen 2002; Gibbons 2002; Padmanabhan 2002; Bagla et al. 2003; Abramo and Finelli 2003; Aguirregabiria and Lazkoz 2004; Guo and Zhang 2004; Copeland et al. 2005), Chaplygin gas model (Kamenshchik et al. 2001; Jackiw 2000; Bilic et al. 2002; Bento et al. 2002) etc. Though the dark energy models mentioned above have field theoretic support and have been introduced to explain cosmological evolution, none of the existing dark energy models is fully satisfactory. Firstly, for a viable scalar-tensor cosmological model the scalar mode has to obey the Chameleon mechanism (Khoury and Weltman 2004; Brax et al. 2004; Gubser and Khoury 2005; Tsujikawa et al. 2009; Ito and Nojiri 2009; Brax et al. 2010). Secondly, these types of exotic scalar fields are presently beyond any all possible scopes to be detected experimentally.

## 6.2 Modified theory of gravity

We have realized that though Einstein gravity is well tested in the solar system it cannot explain the phenomena in very high as well as low curvature regions. Modification in Einstein's theory is therefore another strong possibility to accommodate the phenomena of early universe as well as the phenomena in cosmological scale at late time universe. An alternative approach (Dvali et al. 2000; Carroll et al. 2006; Dvali and Turner 2003; Vollick 2003; Flanagan 2004; Vollick 2004; Soussa and Woodard 2004) is a phenomenological modification of Einstein gravity to obtain an effective contribution of dark energy. The geometrical modifications can arise from quantum effects such as higher curvature corrections to the Einstein Hilbert action. This approach is known as modified theory of gravity. Popularly, such known theories correspond to $F(R), F(G), F(G, R)$ models, where, $G = R_{abcd}R^{abcd} - 4R_{ab}R^{ab} + R^2$ stands for Gauss-Bonnet term. For detailed discussion, one may consult two important review articles (Nojiri and Odintsov 2011; Capozziello and Laurentis 2011).

## 6.3 Late-time Cosmology following particle creation

In a nearly flat R-W model ($k \approx 1$), it is indeed possible that particles might have been created at the expense of gravitational field that we are discussing, or due to some electromagnetic effect (Haouat and Chekireb 2011, 2012), although may be at a very slow rate. Recently, Lima et al. (LSS) (Lima et al. 2008) have developed a late time model universe taking into account particle creation phenomena in the matter dominated era.

In LSS model (Lima et al. 2008), the Friedmann equation, taking into account the created matter, baryonic matter and radiation, reads

$$\left(\frac{H}{H_0}\right)^2 = \Omega_r(1+z)^4 + \Omega_B(1+z)^3 + \frac{\rho_{cm}}{\rho_c} \tag{29}$$

where, $\rho_c$ is the present value of critical density. Straight forward calculation yields

$$\frac{\rho_{cm}}{\rho_c} = \Omega_{cm}(1+z)^3 \exp\left(-\int_t^{t_0} \Gamma dt'\right) \tag{30}$$

where, $\Omega_{cm}$ is the density parameter corresponding to the created matter and $\Gamma$ is the creation parameter mentioned earlier. So, the Friedmann equation finally reads

$$\left(\frac{H}{H_0}\right)^2 = \Omega_r(1+z)^4 + \Omega_B(1+z)^3 + \Omega_{cm}(1+z)^3 e^{\left(-\int_t^{t_0} \Gamma dt'\right)} \tag{31}$$

Now, under the assumption, $\Gamma = 3\beta H + 3\gamma H_0$, where β and γ are constants and $H_0$ is the present Hubble parameter, LSS (Lima et al. 2008) obtained a solution of the scale factor in the form

$$a(t) = a_0 \left[\frac{1-\gamma-\beta}{\gamma}\left(e^{\frac{3\gamma H_0 t}{2}} - 1\right)\right]^{\frac{2}{3(1-\beta)}} \tag{32}$$

which admits the observed transition from early deceleration to late time acceleration. In later investigations (Steigman et al. 2009; Debnath and Sanyal 2011), this model was found to produce a clear conflict between SNIa data at low redshift and the WMAP data constraint on the matter-radiation equality $z_{eq} = 3141 \pm 157$, occurred at the high redshift limit of the observed integrated Sachs–Wolfe (ISW) effect. More precisely, Debnath and Sanyal (Debnath and Sanyal 2011) observed that although this model fits SNIa data to some extent, yields $z_{eq} = 1798^{+536}_{-552}$, instead. Therefore the model does not fit with the WMAP data constraint on the matter-radiation equality $z_{eq} = 3141 \pm 157$, which occurred at the high redshift limit of the observed ISW effect. This contradiction was alleviated by Debnath and Sanyal (Debnath and Sanyal 2011), considering the existence of 26% of primeval matter in the form of baryons (4%) and CDM (22%), that was created in the very early universe and which was responsible for inflation. This amount of CDM created in the very early universe, now behaves as pressure-less dust and has been redshifted like baryons. If we now add the corresponding density parameter, $\Omega_{CDM}$, associated with the CDM created in the very early universe, then the Friedmann equation reads

$$\left(\frac{H}{H_0}\right)^2 = \Omega_r(1+z)^4 + \Omega_m(1+z)^3 + \Omega_{cm}(1+z)^3 e^{\left(-\int_t^{t_0} \Gamma dt'\right)} \tag{33}$$

where, $\Omega_m = \Omega_B + \Omega_{CDM}$. Debnath and Sanyal (Debnath and Sanyal 2011) presented a realistic model by choosing the scale factor judiciously, such that particle creation could start again in the matter-dominated era, instead of choosing the creation parameter $\Gamma$ arbitrarily (Lima et al. 2008). Here, we briefly discussed the model.

A scale factor associated with the so-called intermediate inflation (Barrow 1990; Barrow and Saich 1990), viz $a = a_0 \exp[At^f]$, $a_0$ being a constant was chosen for the purpose. A solution for $A > 0$ and $0 < f < 1$ was shown to lead to late time acceleration (Sanyal 2007, 2008, 2009) in different models. In view of this scale factor, the redshift parameter $z$ is found as

$$1 + z = \frac{a(t_0)}{a(t)} = \exp[A(t_0^f - t^f)] \tag{34}$$

where, $t_0$ is the present time. Hence, the Hubble parameter takes the following form:

$$H = \frac{\dot{a}}{a} = \frac{Af}{t^{(1-f)}} = \frac{Af}{\left[t_0^f - \frac{\ln(1+z)}{A}\right]^{\frac{1-f}{f}}} \tag{35}$$

The form of creation parameter $\Gamma$ is found as

$$\Gamma = 3H + \frac{2\dot{H}}{H} = 3H - 2(1-f)\left(\frac{H}{Af}\right)^{\frac{1}{1-f}} \tag{36}$$

which is clearly different from the β–γ model (Lima et al. 2008). The most important difference is that the creation rate $\Gamma$ here starts developing only when the Hubble parameter

$$H \geq \left[\left(\frac{3}{2(1-f)}\right)^{(1-f)} Af\right]^{\frac{1}{f}} \tag{37}$$

since, $\Gamma < 0$ is not allowed by the second law of thermodynamics. The creation pressure and the creation matter density are now found as

$$8\pi G p_{cm} = -\Gamma H; \qquad 8\pi G \rho_{cm} = 3H^2 - 8\pi G \rho_m \tag{38}$$

where, $8\pi G \rho_m = 8\pi G \rho_{m0}(1+z)^3 = 3H_0^2 \Omega_m (1+z)^3$, in which $\rho_{m0}$ and $\Omega_m$ are the present matter density and the matter density parameter respectively. One can find the effective state parameter and also the state parameter of the created matter as

$$\omega_e = -\frac{2\dot{H} + 3H^2}{3H^2} = -1 + \frac{2}{3}\left(\frac{1-f}{Aft^f}\right); \qquad \omega_{cm} = -\frac{2\dot{H} + 3H^2}{3H^2 - 8\pi G \rho_m} \tag{39}$$

So, the model is parametrized by the two parameters $A$ and $f$. To fit the observed data, the authors (Debnath and Sanyal 2011) kept $0.96 \leq H_0 t_0 \leq 1$ and $0.67 \leq h \left(= \frac{9.78}{H_0^{-1}} Gyr^{-1}\right) \leq 0.7$, at par with the HST project (Freedman 2001). The model was tested by choosing $A$ and $f$ from a wide range of values between $0.08 \leq A \leq 25$ and $0.03 \leq f \leq 0.99$. The Luminosity-distance versus redshift curve (figure 3) fits perfectly with observation, for large $A$ and small $f$ and vice versa.

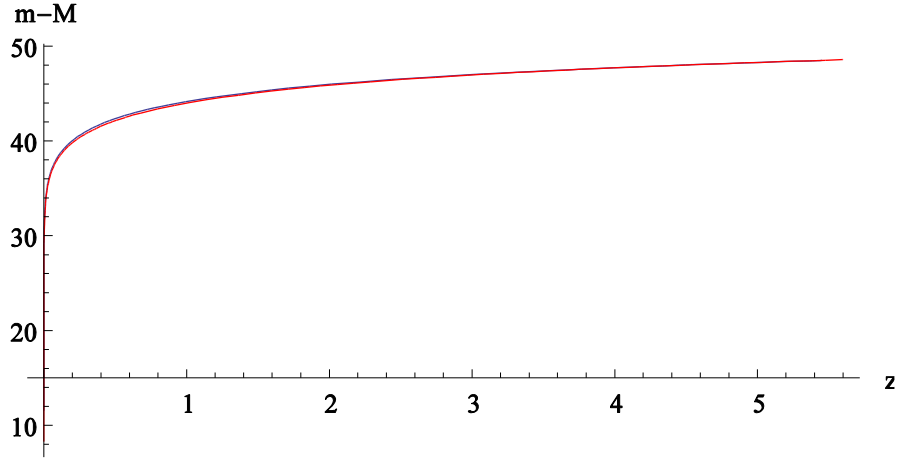

**Figure.3 Distance modulus ($M - m$) versus redshift $z$ plot of the present model (blue), shows perfect fit with the ΛCDM model (red).**

The authors briefly demonstrated the results so obtained, in table 2 of their article (Debnath and Sanyal 2011). The final result is, with, $z_{eq} = 3300$, $\Omega_B = 4\%$, $\Omega_{CDM} = 22\%$, the amount of dark matter produced in the late stage of cosmic evolution, $\Omega_{cm} = 74\%$. This replaces the issue of dark energy solely by the creation of dark matter.

# 7    Concluding remarks

At the end, we understand that to explain late-time accelerated expansion of the universe, either the energy momentum tensor $T_{\mu\nu}$ has to be modified or the geometry itself. Lot of attempts have been made in this regard. Attempts initiated with the modification of $T_{\mu\nu}$, including one or more exotic scalar fields, even Tachyons. It is important to mention that, most of the inflationary models also require such type of fields. However, till date we have not been able to detect a single scalar field, other than the Higgs, which can't be responsible for late time cosmic acceleration. So attempts to explain cosmic evolution taking into account such fields with exotic potentials appear to be yet another search of 'ether'. On the contrary, modified theory of gravity requires scalar-tensor equivalent form (In Jordan or Einstein's frame) for solar test. However, whether these frames are physically equivalent, is a long standing debate (Gasperini and Veneziano 1993, 1994; Magnano and Sokolowski 1994; Dick 1998; Faraoni and Gunzig 1999; Faraoni et al. 1999; Nojiri and Odintsov 2006; Capozziello et al. 2006; Bhadra et al. 2007; Briscese et al. 2007; Capozziello et al. 2010; Brooker et al. 2016; Banerjee and Majumder 2016; Bahamonde et al. 2016; Sk and Sanyal 2016).

Here, we concentrated on yet another attempt towards modifying $T_{\mu\nu}$, considering particle creation phenomena. This appears to be much logical although cold dark matter (CDM) has also not been detected as yet. The reason that we belief this to be the most powerful candidate is that, one can't avoid CDM in any case. As we know, without CDM, there is presently no explanation to the structure formation. Lot of experiments are carried out presently to detect different components of CDM. We believe that within a decade or so, CDM will be detected. The question that would arise is how much CDM is presently available in the universe? If it is around 20%, then it has been created only in the very early universe. If it is around 96% then it has been created also in the late universe, which would explain late time accelerated expansion without the need of dark energy, or modified theory of gravity.

In view of the above discussions, we strongly believe particle creation phenomenon is the most powerful theory developed so far to explain cosmic evolution and it still needs lot of further attention.

Here, some of the aspects of the cosmological models beyond the standard model have been discussed for both the early and late era. The recent predictions from different cosmological observations have been considered constructing different cosmological models. A number of issues are addressed, however to understand more clearly we have to wait for data from a number of future astronomical and cosmological observatories coming up in the future. A host of experiments are being carried out recently to detect a WIMP, viz., neutralino (a possible candidate for the dark matter). Direct detection of neutralino is being carried out under the Gran Sasso Mountain in Italy, by Italian-Chinese collaboration DAMA (short for Dark MAtter) (Nosengo 2012) taking sodium iodide crystal (a scintillator) as the detector. Besides the direct detection of galactic neutralino in the laboratory, high energy neutrinos from the core of the Sun or of the Earth as a result of neutralino annihilation can be detected in Cherenkov neutrino telescopes. Several neutrino telescopes are currently operational, viz., the Super-Kamiokande detector (Kearns et al. 1999) in Japan, the AMANDA detector (IceCube Collaboration and IPN Collaboration 2008) at the South Pole, ANTARES detector (ANTARES Collaboration 2012) and the NESTOR detector in the Mediterranean (Resvanis 1992). There is also a new GLAST detector (Cheung et al. 2016) with an adequate energy resolution. This may detect gamma-rays and cosmic rays arising from neutralino annihilation in galactic halos in the energy range 10 GeV - 10 TeV.